\documentclass[conference]{IEEEtran}
\usepackage{subfigure}
\usepackage{cite}
\usepackage{citesort}
\usepackage{amsmath,epsfig,bm}
\usepackage{pifont}
\usepackage{graphicx}
\usepackage{amssymb}
\usepackage{amsmath}
\usepackage{cancel}
\usepackage[normalem]{ulem}
\usepackage[dvips]{color}
\usepackage{algorithm}

\DeclareMathOperator*{\argmin}{argmin}

\usepackage{pifont}
\usepackage{graphicx}
\usepackage{amssymb}
\usepackage{amsmath}
\usepackage{cancel}
\usepackage[normalem]{ulem}
\usepackage[dvips]{color}
\usepackage{algorithm}
\usepackage{algorithmic}

\newtheorem{theorem}{\noindent \textbf{Theorem}}

\newcommand{\Hnull}{\mathcal{H}_0}
\newcommand{\Halt}{\mathcal{H}_1}

\newcommand{\Honull}{\mathcal{D}_0}
\newcommand{\Hoalt}{\mathcal{D}_1}

\begin{document}
%%%%%%%%%%%%%%%%%%%%%%%%%%%%%%%%%%%%%%%%%%%%%%%%%
\title{Location Spoofing Detection for VANETs by a Single Base Station in Rician Fading Channels}
%%%%%%%%%%%%%%%%%%%%%%%
\author{
\IEEEauthorblockN{Shihao Yan$^1$, Robert Malaney$^1$}
\IEEEauthorblockA{$^1$School of Electrical Engineering  \& Telecommunications,\\
The University of New South Wales,\\
Sydney, NSW 2052, Australia\\
Emails: shihao.yan@unsw.edu.au;  r.malaney@unsw.edu.au}
\and
\IEEEauthorblockN{Ido Nevat$^2$, Gareth W. Peters$^{3}$}
\IEEEauthorblockA{$^2$Institute for Infocomm Research, A$^{\star}$STAR, Singapore.\\
%$^3$School of Mathematics and Statistics, University of NSW\\
$^3$Department of Statistical Science,\\
University College London, London, UK\\
Emails: ido-nevat@i2r.a-star.edu.sg; gareth.peters@ucl.ac.uk
}}

%%%%%%%%%%%%%%%%%%%%%%%
\vspace{-6cm}

\maketitle

%%%%%%%%%%%%%%%%%%%%%%%%%%%%%%%%%%%%%%%%%%%%%%%%%%%
% ABSTRACT
%%%%%%%%%%%%%%%%%%%%%%%%%%%%%%%%%%%%%%%%%%%%%%%%%%%
\begin{abstract}
In this work we examine the performance of a Location Spoofing Detection System (LSDS) for vehicular networks in the realistic setting of Rician fading channels. In the LSDS, an authorized Base Station (BS) equipped with multiple antennas utilizes  channel observations to identify a malicious  vehicle, also equipped with multiple antennas, that is spoofing its location.
After deriving the optimal transmit power and the optimal directional beamformer of a potentially malicious vehicle, robust theoretical analysis and detailed simulations are conducted in order to determine
the impact of key system parameters on the LSDS performance. Our analysis shows how LSDS performance increases as the Rician $K$-factor of the channel between the BS and legitimate vehicles increases, or as the number of antennas at the BS or legitimate vehicle increases. We also obtain the counter-intuitive result that the malicious vehicle's optimal number of antennas conditioned on its optimal directional beamformer is equal to the legitimate vehicle's number of antennas. The results we provide here are important for the verification of location information reported in IEEE 1609.2 safety messages.
\end{abstract}

\section{Introduction}

 In wireless networks the integrity of location information is of growing importance. As such, the authentication (or verification) of location information has attracted considerable research interest in recent years \cite{malaney2004location,vora2006secure,chen2010detecting,chiang2012secure,yan2014optimal,yan2014signal,yanreview}. In many circumstances the device (client) itself obtains its location information directly (e.g., via GPS),  and the wider network can only achieve the client's location information through requests to the client. In such a context, the client can easily spoof or falsify its claimed location in order to  disrupt some network functionality  (e.g., geographic routing protocols \cite{leinmuller2005influence}, or directional access control protocols \cite{capkun2010integrity}). The adverse effects of
location spoofing can be more severe in Vehicular Ad Hoc Networks (VANETs) due to the possibility of life-threatening  accidents. Less critically, a malicious vehicle could spoof its location in order to seriously disrupt other vehicles \cite{raya2007securing} or to selfishly enhance its own functionality within the network \cite{yu2013detecting}. The integrity of claimed location  in VANETs is therefore important, and motivates the introduction of an LSDS to that scenario. Within IEEE 1609.2  revocation of certificates belonging to malicious vehicles will occur \cite{tang} - an LSDS will form part of the revocation logic.

%%%%%%%%%%%%%%%%%%%%%%%%%%%%

Recently, many location spoofing detection protocols for VANETs have been proposed  (e.g.,\cite{leinmuller2006position,yan2009providing,abumansoor2012a}). In \cite{leinmuller2006position}, the authors proposed an autonomous and cooperative scheme for detecting and mitigating false claimed locations by exploiting specific properties of VANETs, such as high node density and mobility. The authors of \cite{yan2009providing} developed a location spoofing detection algorithm by comparing the claimed location with a neighbor table consisting of other vehicles' identifications and locations.  To overcome the non line-of-sight (LOS)  problem in location verification systems, a cooperative algorithm was proposed in \cite{abumansoor2012a}.
%Since no specific threat model was proposed in the aforementioned works, the detection performances of these protocols were not theoretically analyzed.
Some generic location spoofing detection algorithms (not dedicated to VANETs), were also proposed in recent years (e.g., \cite{chen2010detecting,chiang2012secure,yan2014optimal,yan2014signal}). These algorithms utilize some observations such as Received Signal Strength (RSS), Time of Arrival (TOA), and Angle of Arrival (AOA), and  performance analysis of these algorithms were provided under specific observation models.

However, the following  question has not been explored in the literature. \emph{How does the performance of an LSDS depend on the proportion of the channel which is LOS?}  In the VANET environment  it is highly likely that a vehicle possesses some LOS component towards a BS for the majority of its travel time. As such, answering the above question in the context of VANETs is important, and  forms the thrust of the work presented here.
In order to investigate the above question, we consider Rician fading channels in which the Rician $K$-factor is defined as the ratio between the power of the LOS component and the power of other scattered components. We utilize the complex signals measured by an authorized multiple-antenna BS to verify a claimed location of a vehicle that is equipped with multiple antennas, and infer whether the vehicle is \emph{legitimate} (reporting its true location) or \emph{malicious} (spoofing its claimed location). Adopting a practical threat model, in which the on-road malicious vehicle keeps some minimum distance away from its claimed location, we analyze the performance of our LSDS. In order to guarantee  fairness, we also determine the optimal transmit power and the optimal directional beamformer for the malicious vehicle to minimize the detection rate. Our analysis demonstrates that our LSDS works well even when the Rician $K$-factor is low (e.g., $-3$dB), and that detection performance  increases as the Rician $K$-factor of the channel between the legitimate vehicle and the BS  increases. We also obtain a counter-intuitive observation, that the malicious vehicle can minimize the detection rate by setting its number of antennas equal to the legitimate vehicle's number of antennas when it adopts the optimal directional beamformer. This is counter-intuitive since \emph{a priori} one would have thought it would be optimal to set the number of antennas as large as possible.

\section{System Model}\label{System Model}

\subsection{System Assumptions}

\begin{figure}[!t]
    \begin{center}
   {\includegraphics[width=3.1in]{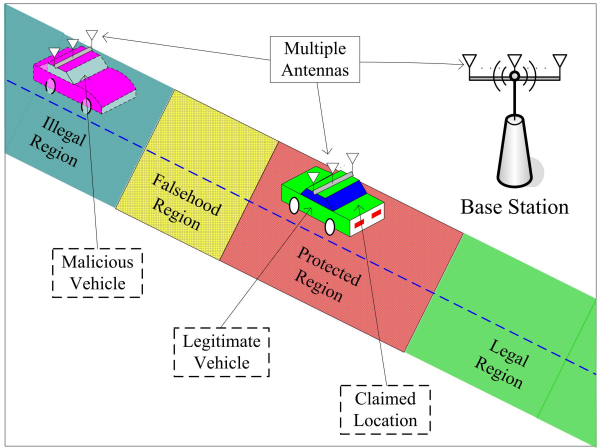}}
    \caption{Illustration of the VANETs application scenario of interest.}\label{fig:scenario}
    \end{center}
\end{figure}

\begin{figure}[!t]
    \begin{center}
   {\includegraphics[width=3in]{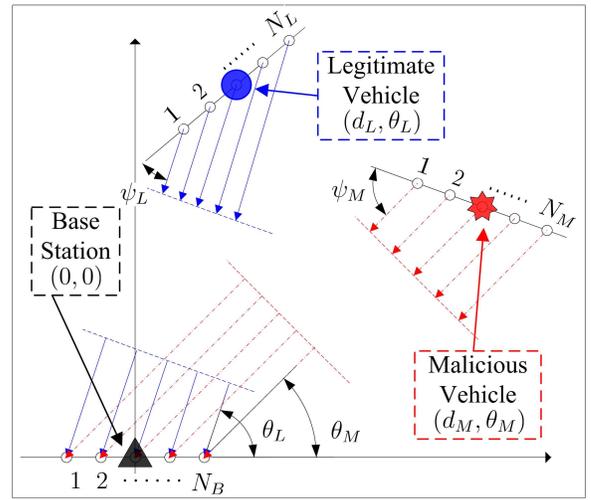}}
    \caption{Illustration of the orientations of the three ULA antennas and the geometry of the BS, legitimate vehicle, and malicious vehicle.}\label{fig:system}
    \end{center}
\end{figure}

The VANETs application scenario of interest is illustrated in Fig.~\ref{fig:scenario}, where the BS, the legitimate vehicle, and the malicious vehicle are each equipped with  a uniform linear array (ULA) with $N_B$, $N_L$, and $N_M$ elements, respectively. In this figure, the ``Protected Region'' is the area where a vehicle (legitimate or malicious) claims to be. The BS is to verify whether the vehicle is indeed at his claimed location or not based on wireless channel observations. If the vehicle passes such a verification, a specific action will follow in the ``Legal Region'' (e.g., a traffic light turns green). The ``Falsehood Region'' indicates the minimum distance between the claimed location and the malicious vehicle's location. The malicious vehicle is inside the ``Illegal Region'' while it claims that it is inside the ``Protected Region'' in order to bring some selfish benefits (e.g., it keeps the traffic light green in advance for itself).
We adopt the polar coordinate system  as shown in Fig.~\ref{fig:system}, where the location of the BS is selected as the origin, the legitimate vehicle's location is denoted as $\mathbf{x}_L = (d_L, \theta_L)$, and the malicious vehicle's location is denoted as $\mathbf{x}_M = (d_M, \theta_M)$. We assume that the claimed location is the same as the legitimate vehicle's location (i.e., $\mathbf{x}_L$ is also the claimed location of the legitimate or malicious vehicle). We adopt a practical threat model, in which the distance between $\mathbf{x}_M$ and the malicious vehicle's claimed location $\mathbf{x}_L$ is larger than a predetermined threshold $r_m$, i.e., $|\mathbf{x}_M - \mathbf{x}_L| \geq r_m$. The orientation of the  BS ULA is aligned with the $x$-axis, which is publicly known. The orientations of the ULAs of the legitimate and malicious vehicles are under the control of the legitimate and malicious vehicles, respectively, i.e., the angles $\psi_L$ and $\psi_M$ as shown in Fig.~\ref{fig:system} are under the control of the legitimate and malicious vehicles, respectively.

\subsection{Channel Model}

With no loss of generality, we assume the legitimate channel (legitimate vehicle-to-BS) and the malicious channel (malicious vehicle-to-BS) are subject to Rician fading. Then, the $N_B \times N_L$ legitimate channel matrix is given by
\begin{align}\label{h_definition}
\mathbf{H} = \sqrt{\frac{K_L}{1+K_L}} \mathbf{H}_o + \sqrt{\frac{1}{1+K_L}} \mathbf{H}_r,
\end{align}
where $K_L$ is the Rician $K$-factor of the legitimate channel, $\mathbf{H}_o$ is the LOS component of the matrix, and $\mathbf{H}_r$ is the scattered component of the matrix. The entries of $\mathbf{H}_r$ are independent and identically distributed (i.i.d) circularly-symmetric complex Gaussian random variables with zero mean and unit variance. Denoting $\rho_B$ as the space between two antenna elements of the ULA at the BS, $\mathbf{H}_o$ is given by
$\mathbf{H}_o = \mathbf{r}_{L}^T  \mathbf{t}_L$,
where $\mathbf{r}_L$ and $\mathbf{t}_L $ are defined as
\begin{align}
\mathbf{r}_{L} &= \left[1,\cdots,\exp(j (N_B -1)\tau_B \cos \theta_L)\right],\label{r_L_definition} \\
\mathbf{t}_L  &= \left[1,\cdots,\exp(j (N_L -1)\tau_L \cos \psi_L)\right], \label{t_L_definition}
\end{align}
and $^T$ denotes the transpose operation.
In \eqref{r_L_definition} and \eqref{t_L_definition}, we have $\tau_B = 2 \pi f_0 \rho_B /c$ and $\tau_L = 2 \pi f_0 \rho_L /c$, where $f_0$ is the carrier frequency, $c$ is the speed of propagation of the plane wave, and $\rho_L$ is the space between two antenna elements of the ULA at the legitimate vehicle.

The $N_B \times N_M$ malicious channel matrix is given by
\begin{align}\label{g_definition}
\mathbf{G} = \sqrt{\frac{K_M}{1+K_M}} \mathbf{G}_o + \sqrt{\frac{1}{1+K_M}} \mathbf{G}_r,
\end{align}
where $K_M$ is the Rician $K$-factor of the malicious channel, $\mathbf{G}_o$ is the LOS component of the matrix. $\mathbf{G}_r$ is the scattered component of the matrix and is a matrix with i.i.d circularly-symmetric complex Gaussian random variables with zero mean and unit variance.  $\mathbf{G}_o$ can be written as
$\mathbf{G}_o = \mathbf{r}_M^T \mathbf{t}_M$,
where $\mathbf{r}_M$ and $\mathbf{t}_M$ are given by
\begin{align}
\mathbf{r}_{M} &= \left[1,\cdots,\exp(j (N_B -1)\tau_B \cos \theta_M)\right],\label{r_M_definition}
\end{align}
\begin{align}
\mathbf{t}_M &= \left[1,\cdots,\exp(j (N_M -1)\tau_M \cos \psi_M)\right], \label{t_M_definition}
\end{align}
In \eqref{t_M_definition}, we have $\tau_M = 2 \pi f_0 \rho_M /c$, where $\rho_M$ is the space between two antenna elements of the ULA at the malicious vehicle.

\subsection{Observation Model}

Throughout this work we denote the null hypothesis where the vehicle is legitimate as $\Hnull$ and denote the alternative hypothesis where the vehicle is malicious as $\Halt$. The $N_B \times 1$ complex observation vector received from the legitimate vehicle (under $\Hnull$) is given by
\begin{align}\label{y_received0}
\mathbf{y} = \sqrt{P_L \mathrm{g}(d_L)}\mathbf{H}\mathbf{b} s + \mathbf{n}_L,
\end{align}
where $P_L$ is the transmit power of the legitimate vehicle, $\mathrm{g}(d_L)$ is the path loss gain given by $\mathrm{g}(d_L) = \left(c/4\pi f_0 d_0\right)^2({d_0}/{d_L})^{\eta}$, $d_0$ is a reference distance, $\eta$ is the path loss exponent, $\mathbf{b}$ is the beamformer adopted by the legitimate vehicle which satisfies $\|\mathbf{b}\| = 1$, $s$ is the publicly known pilot symbol (without loss of generality we assume $s = 1$), and $\mathbf{n}_L$ is the additive white Gaussian noise vector, of which the entries are i.i.d circularly-symmetric complex Gaussian random variables with zero mean and variance $\sigma_L^2$.
We note that $\mathbf{b}$ and $P_L$ are under the control of the legitimate vehicle. We assume that the legitimate vehicle cooperates with the BS to facilitate the verification procedure. To this end, the legitimate vehicle sets $\mathbf{b} = \mathbf{t}_L ^{\dag}/\|\mathbf{t}_L \|$ to maximize $|\mathbf{t}_L  \mathbf{b}|$, where $^{\dag}$ denotes the conjugate transpose operation. In addition, the legitimate vehicle sets its transmit power to that  required by the BS (we assume $P_L$ is publicly known).
As per \eqref{y_received0}, the likelihood function of $\mathbf{y}$ conditioned on a known $s$ under $\Hnull$  is
\begin{align}\label{pdf_y0}
f(\mathbf{y}|\Hnull) \!=\! \frac{1}{\pi^{N_B}\det(\mathbf{R}_0)} \exp\left[\!-\!(\mathbf{y}\!\!-\!\!\mathbf{m}_0)^{\dag}\mathbf{R}_0^{\!-\!1}(\mathbf{y}\!\!-\!\!\mathbf{m}_0)\right],
\end{align}
where $\mathbf{m}_0$ and $\mathbf{R}_0$ are the mean vector and covariance matrix of $\mathbf{y}$ under $\Hnull$, respectively, which are given by
\begin{align}
\mathbf{m}_0 &= \sqrt{\frac{P_L \mathrm{g}(d_L) K_L N_B}{1+K_L}}\mathbf{r}_L^T ,\\
\mathbf{R}_0 &= \left(\frac{P_L \mathrm{g}(d_L)}{K_L + 1} + \sigma_L^2\right) \mathbf{I}_{N_B}.\label{R0_definition}
\end{align}

Likewise, the complex observation vector received from the malicious vehicle (under $\Halt$) is given by
\begin{align}\label{y_received1}
\mathbf{y} &= \sqrt{P_M \mathrm{g}(d_M)}\mathbf{G}\mathbf{p} s + \mathbf{n}_M,
\end{align}
where $P_M$ is the transmit power of the malicious vehicle, $\mathrm{g}(d_M)$ is the path loss gain given by $\mathrm{g}(d_M) = \left(c/4\pi f_0 d_0\right)^2({d_0}/{d_M})^{\eta}$, $\mathbf{p}$ is the beamformer adopted by the malicious vehicle which satisfies $\|\mathbf{p}\| = 1$, and $\mathbf{n}_M$ is the additive white Gaussian noise vector, of which the entries are i.i.d circularly-symmetric complex Gaussian random variables with zero mean and variance $\sigma_M^2$. As per \eqref{y_received1}, the likelihood function of $\mathbf{y}$ under $\Halt$ for given $\mathbf{x}_M$, $P_M$, and $\mathbf{p}$ is
\begin{align}\label{pdf_y1}
f(\mathbf{y}|&\mathbf{x}_M, P_M, \mathbf{p}, \Halt)\notag\\
&= \frac{1}{\pi^{N_B}\det(\mathbf{R}_1)} \exp\left[\!-\!(\mathbf{y}\!-\!\mathbf{m}_1)^{\dag}\mathbf{R}_1^{\!-\!1}(\mathbf{y}\!-\!\mathbf{m}_1)\right],
\end{align}
where $\mathbf{m}_1$ and $\mathbf{R}_1$ are the mean vector and covariance matrix of $\mathbf{y}$ under $\Halt$, respectively, which are given by
\begin{align}
\mathbf{m}_1 &= \sqrt{\frac{P_M \mathrm{g}(d_M) K_M}{1+K_M}}\mathbf{G}_o \mathbf{p},\\
\mathbf{R}_1 &= \left(\frac{P_M \mathrm{g}(d_M)}{K_M + 1} + \sigma_M^2\right) \mathbf{I}_{N_B}.
\end{align}
We note that $\mathbf{x}_M$, $P_M$, and $\mathbf{p}$ are under the control of the malicious vehicle. We will discuss in the next section how  the malicious vehicle sets these parameters so as to minimize the detection rate.

\section{Location Spoofing Detection System}

In this section, we first present the decision rule adopted in our LSDS. We then discuss the attack strategy of the malicious vehicle (e.g., how to set $\mathbf{x}_M$, $P_M$, and $\mathbf{p}$) in order to minimize the detection rate. Finally, we analyze the detection performance of our LSDS based on the malicious vehicle's attack strategy.

\subsection{Decision Rule of the LSDS}

We adopt the Likelihood Ratio Test (LRT) as the decision rule of our LSDS. This is due to the fact that the LRT achieves the highest detection rate (the probability to detect a malicious vehicle) for any given false positive rate (the probability to detect a legitimate vehicle as malicious) \cite{neyman1933problem}. The LRT decision rule is given by
\begin{equation}\label{arbitrary}
\Lambda\left(\mathbf{y}\right) \triangleq \frac{f\left(\mathbf{y}|\mathbf{x}_M, P_M, \mathbf{p}, \Halt\right)}{f\left(\mathbf{y}|\Hnull\right)} \begin{array}{c}
\overset{\Hoalt}{\geq} \\
\underset{\Honull}{<}
\end{array}%
\lambda,
\end{equation}
where $\Lambda\left(\mathbf{y}\right)$ is the likelihood ratio of $\mathbf{y}$, $\lambda$ is the threshold corresponding to $\Lambda\left(\mathbf{y}\right)$, and $\Honull$ and $\Hoalt$ are the binary decisions that infer whether the vehicle is legitimate or malicious, respectively. Given the decision rule in \eqref{arbitrary}, the false positive and detection rates of an LSDS are functions of $\lambda$. The specific value of $\lambda$ can be set through predetermining a false positive rate, minimizing the Bayesian average cost, or maximizing the mutual information between the system input and output \cite{yan2014optimal}. In this work, we adopt the false positive rate, $\Pr\left(\Lambda\left(\mathbf{y}\right) > \lambda|\Hnull\right)$, and detection rate, $\Pr\left(\Lambda\left(\mathbf{y}\right) > \lambda|\Halt\right)$, as the core performance metrics for our LSDS. In addition, we adopt the minimum total error as the unique performance metric in order to investigate the impact of key system parameters on the performance of our LSDS.

\subsection{Attack Strategy of the Malicious Vehicle}

We assume the malicious vehicle knows all the information known by the BS or the legitimate vehicle. We first discuss how does the malicious vehicle set its true location $\mathbf{x}_M$. Since there is only one BS in our LSDS, the difference between $d_L$ and $d_M$ can be eliminated by the malicious vehicle through adjusting its transmit power $P_M$. This is the reason why a single BS cannot detect location spoofing attacks based on the RSS of a channel. As such, we assume the malicious vehicle sets $\mathbf{x}_M$ by minimizing the difference between $\theta_M$ and $\theta_L$ under the constraint $|\mathbf{x}_M - \mathbf{x}_L| \geq r_m$. Then, the adopted value of $\mathbf{x}_M$ can be obtained through
\begin{align}
\mathbf{x}_M^{\ast} \triangleq (d_M^{\ast}, \theta_M^{\ast}) = \argmin_{|\mathbf{x}_M - \mathbf{x}_L| \geq r_m} |\theta_M -\theta_L|.
\end{align}
Given the application scenario of interest as shown in Fig.~\ref{fig:scenario}, we assume that $\mathbf{x}_M^{\ast}$ is known to the BS.
The average signal-to-noise ratio (SNR) of a channel can be readily estimated. As such, we assume that the malicious vehicle adjusts its transmit power to make sure that the average SNR of the malicious channel is the same as that of the legitimate channel, i.e., $\overline{\gamma}_M = \overline{\gamma}_L$, where $\overline{\gamma}_L = P_L \mathrm{g}(d_L)/\sigma_L^2$ and $\overline{\gamma}_M = P_M \mathrm{g}(d_M)/\sigma_M^2$. Therefore, the transmit power of the malicious vehicle conditioned on $\mathbf{x}_M$ is given by
\begin{align}
P_M^{\ast}(\mathbf{x}_M) = \frac{P_L \mathrm{g}(d_L) \sigma_M^2}{\mathrm{g}(d_M) \sigma_L^2}.
\end{align}

We next discuss how does the malicious vehicle sets its beamformer $\mathbf{p}$, which is the key vector controlled by the malicious vehicle. The Kullback-Leibler (KL) divergence from $f\left(\mathbf{y}|\mathbf{x}_M, P_M, \mathbf{p}, \Halt\right)$ to $f\left(\mathbf{y}|\Hnull\right)$ is defined as
\begin{align}
&D_{KL}\left(f\left(\mathbf{y}|\mathbf{x}_M, P_M, \mathbf{p}, \Halt\right)||f\left(\mathbf{y}|\Hnull\right)\right) \notag \\
&~~~~~~~~~~~~~~~~= \int \ln  \Lambda (\mathbf{y}) f\left(\mathbf{y}|\mathbf{x}_M, P_M, \mathbf{p}, \Halt\right) d \mathbf{y}. \label{KL_definition}
\end{align}
As per \eqref{KL_definition}, we know that the KL divergence is also the expected log likelihood ratio when the alternative hypothesis is true. Based on \eqref{arbitrary}, we also know that the larger the KL divergence, the more evidence we have for the alternative hypothesis \cite{eguchi2006interpreting}. As such, the malicious vehicle is to minimize the KL divergence presented in \eqref{KL_definition} in order to minimize the detection rate. Substituting \eqref{pdf_y0} and \eqref{pdf_y1} into \eqref{KL_definition}, we have
\begin{align}\label{KL_result}
D_{KL}&\left(f\left(\mathbf{y}|\mathbf{x}_M, P_M, \mathbf{p}, \Halt\right)||f\left(\mathbf{y}|\Hnull\right)\right)
= \text{tr}(\mathbf{R}_0^{-1}\mathbf{R}_1) \!-\!N_B\notag \\
&\!-\! \ln \left(\frac{\det \mathbf{R}_1}{\det \mathbf{R}_0}\right)
+ \underbrace{(\mathbf{m}_0 \!-\! \mathbf{m}_1)^{\dag}\mathbf{R}_0^{-1}(\mathbf{m}_0 \!-\! \mathbf{m}_1)}_{f_D(\mathbf{p})}.
\end{align}
As per \eqref{KL_result}, we know that only the term $f_D\left(\mathbf{p}\right)$ is a function of $\mathbf{p}$. As such, the optimal $\mathbf{p}$ is the one that minimizes $f_D\left(\mathbf{p}\right)$.
Given the format of $\mathbf{R}_0$ presented in \eqref{R0_definition}, we can see that $f_B\left(\mathbf{p}\right)$ is minimized when $\|\mathbf{m}_0 - \mathbf{m}_1\|$ is minimized.
A constraint on our solution is that we
assume  the malicious vehicle adopts a directional beamformer,
the direction of which is chosen (see below) so as to minimize detection. The rationale for this assumption is that it allows the attacker to optimize his solution based on only one parameter (allowing rapid in-the-field decision making), and allows for a clarity of exposition. The format of our directional beamformer $\mathbf{p}$ is given by
\begin{align}\label{p_format}
\mathbf{p} &= \frac{1}{\sqrt{N_M}}\left[1,\cdots,\exp(j (N_M -1)\tau_M \cos \varphi)\right]^T,
\end{align}
where $\varphi$ is the beamforming direction. Then, the optimal beamforming direction $\varphi$ (the value of $\varphi$ that minimizes the detection rate) conditioned on $\mathbf{x}_M$ and $P_M$ can be obtained through
\begin{align}
\varphi^{\ast} (\mathbf{x}_M, P_M)
= \argmin_{\varphi \in [0,\pi]}\left\|\mathbf{m}_0- \mathbf{K} \mathbf{p}\right\|,\label{opt_p_state}
\end{align}
where $\mathbf{K} = \sqrt{{P_M \mathrm{g}(d_M) K_M}/(1+K_M)}\mathbf{G}_o$. Substituting $\varphi^{\ast} (\mathbf{x}_M, P_M)$ into \eqref{p_format}, we obtain the optimal directional beamformer of the malicious vehicle, denoted as $\mathbf{p}^{\ast}(\mathbf{x}_M, P_M)$. We note that $\mathbf{p}^{\ast}(\mathbf{x}_M, P_M)$ may not be the globally optimal beamformer (only near-optimal) for the malicious vehicle due to the imposition of the one-parameter solution ($\varphi$) of
\eqref{p_format} in obtaining $\mathbf{p}^{\ast}(\mathbf{x}_M, P_M)$.

\subsection{Detection Performance of the LSDS}
Without loss of generality, we first analyze the detection performance of our LSDS for a general $\mathbf{x}_M$. Obviously, the malicious vehicle will optimize its transmit power $P_M$ and its beamformer $\mathbf{p}$ for a given $\mathbf{x}_M$. As such, the following analysis is for  $P_M = P_M^{\ast}(\mathbf{x}_M)$, and $\mathbf{p} = \mathbf{p}^{\ast}\left[\mathbf{x}_M, P_M^{\ast}(\mathbf{x}_M)\right]$. In order to derive the false positive and detection rates in closed-form expressions, we further assume $\sigma_L^2 = \sigma_M^2$ and $K_L =K_M$ such that $\mathbf{R}_0 = \mathbf{R}_1$. We would like to highlight that these assumptions are practical since the malicious vehicle will not be very far from its claimed location in order to keep a low detection rate. Also, as we show later the detection rate is minimized when $K_L =K_M$, i.e., $K_L =K_M$ is the best case for the malicious vehicle. We will also assume the system knows $K_L$, through a predetermined measurement campaign in the vicinity of the BS. In principle, knowledge of $K_L$ could be replaced by a pdf which is then encapsulated within the LSDS decision framework. Substituting \eqref{pdf_y0} and \eqref{pdf_y1} into \eqref{arbitrary}, the LRT decision rule can be rewritten as
\begin{align}\label{decision_rule}
\mathbb{T}(\mathbf{y}) \begin{array}{c}
\overset{\Hoalt}{\geq} \\
\underset{\Honull}{<}
\end{array}%
\Gamma,
\end{align}
where $\mathbb{T}(\mathbf{y})$ is the test statistic given by
\begin{align}\label{test_statistic}
\mathbb{T}(\mathbf{y}) = 2 \text{Re}\{[\mathbf{m}_1^{\ast}(\mathbf{x}_M) - \mathbf{m}_0]^{\dag}\mathbf{R}_0^{-1}\mathbf{y}\},
\end{align}
$\Gamma$ is the threshold for $\mathbb{T}(\mathbf{y})$ given by
\begin{align}
\Gamma \!=\! \ln \lambda + \text{Re}\{[\mathbf{m}_1^{\ast}(\mathbf{x}_M) \!-\! \mathbf{m}_0]^{\dag}\mathbf{R}_0^{-1}[\mathbf{m}_1^{\ast}(\mathbf{x}_M) \!+\! \mathbf{m}_0]\},
\end{align}
$\mathbf{m}_1^{\ast}(\mathbf{x}_M)$ is given by
\begin{align}
\mathbf{m}_1^{\ast}(\mathbf{x}_M) = \sqrt{\frac{P_L \mathrm{g}(d_L) K_L}{1+K_L}}\mathbf{G}_o \mathbf{p}^{\ast}\left[\mathbf{x}_M, P_M^{\ast}(\mathbf{x}_M)\right],
\end{align}
and $\text{Re}\{\}$ denotes the real part of a complex number.
Then, we derive the false positive rate, $\alpha(\lambda,\mathbf{x}_M)$, and the detection rate, $\beta(\lambda,\mathbf{x}_M)$, of the LSDS in the following theorem.
\begin{theorem}
The false positive and detection rates of the LSDS for a given $\mathbf{x}_M$ are given by
\begin{align}
\alpha(\lambda,\mathbf{x}_M)
%&\!=\! \mathcal{Q}\left\{\frac{\Gamma - 2\text{Re}\{[\mathbf{m}_1^{\ast}(\mathbf{x}_M) - \mathbf{m}_0]^{\dag}\mathbf{R}_0^{-1}\mathbf{m}_0\}}{\sqrt{2[\mathbf{m}_1^{\ast}(\mathbf{x}_M) \!-\! \mathbf{m}_0]^{\dag}\mathbf{R}_0^{\!-\!1}[\mathbf{m}_1^{\ast}(\mathbf{x}_M) \!-\! \mathbf{m}_0]}}\right\}\notag\\
&\!=\! \mathcal{Q}\left\{\frac{\ln \lambda + [\mathbf{m}_1^{\ast}(\mathbf{x}_M) \!-\! \mathbf{m}_0]^{\dag}\mathbf{R}_0^{\!-\!1}[\mathbf{m}_1^{\ast}(\mathbf{x}_M) \!-\! \mathbf{m}_0]}{\sqrt{2[\mathbf{m}_1^{\ast}(\mathbf{x}_M) \!-\! \mathbf{m}_0]^{\dag}\mathbf{R}_0^{\!-\!1}[\mathbf{m}_1^{\ast}(\mathbf{x}_M) \!-\! \mathbf{m}_0]}}\right\},\label{alpha_r}
\end{align}
%====================================================================
\begin{align}
\beta(\lambda,\mathbf{x}_M)
%&\!=\! \mathcal{Q}\left\{\frac{\Gamma - 2\text{Re}\{[\mathbf{m}_1^{\ast}(\mathbf{x}_M) - \mathbf{m}_0]^{\dag}\mathbf{R}_0^{-1}\mathbf{m}_1^{\ast}(\mathbf{x}_M)\}}{\sqrt{2[\mathbf{m}_1^{\ast}(\mathbf{x}_M) \!-\! \mathbf{m}_0]^{\dag}\mathbf{R}_0^{\!-\!1}[\mathbf{m}_1^{\ast}(\mathbf{x}_M) \!-\! \mathbf{m}_0]}}\right\}\notag\\
&\!=\! \mathcal{Q}\left\{\frac{\ln \lambda - [\mathbf{m}_1^{\ast}(\mathbf{x}_M) \!-\! \mathbf{m}_0]^{\dag}\mathbf{R}_0^{\!-\!1}[\mathbf{m}_1^{\ast}(\mathbf{x}_M) \!-\! \mathbf{m}_0]}{\sqrt{2[\mathbf{m}_1^{\ast}(\mathbf{x}_M) \!-\! \mathbf{m}_0]^{\dag}\mathbf{R}_0^{\!-\!1}[\mathbf{m}_1^{\ast}(\mathbf{x}_M) \!-\! \mathbf{m}_0]}}\right\},\label{beta_r}
\end{align}
where $\mathcal{Q}(x) = \frac{1}{2 \pi}\int_x^{\infty}\exp\left(-\frac{t^2}{2}\right)d t$.
\end{theorem}
\begin{IEEEproof}
As per \eqref{test_statistic}, we derive the distributions of the test statistic $\mathbb{T}(\mathbf{y})$ under $\Hnull$ and $\Halt$ as follows
\begin{align}
\mathbb{T}(\mathbf{y})|\Hnull &\sim \mathcal{N}\bigg(2\text{Re}\{[\mathbf{m}_1(\mathbf{x}_M) \!-\! \mathbf{m}_0]^{\dag}\mathbf{R}_0^{\!-\!1}\mathbf{m}_0\}, \bigg.\notag \\
&~~~~~~\bigg.2[\mathbf{m}_1^{\ast}(\mathbf{x}_M) \!-\! \mathbf{m}_0]^{\dag}\mathbf{R}_0^{\!-\!1}[\mathbf{m}_1^{\ast}(\mathbf{x}_M) \!-\! \mathbf{m}_0]\bigg),\label{d_t0}\\
\mathbb{T}(\mathbf{y})|\Halt &\sim \mathcal{N}\bigg(2\text{Re}\{[\mathbf{m}_1(\mathbf{x}_M) \!-\! \mathbf{m}_0]^{\dag}\mathbf{R}_0^{\!-\!1}\mathbf{m}_1(\mathbf{x}_M)\}, \bigg.\notag \\
&~~~~~~\bigg.2[\mathbf{m}_1^{\ast}(\mathbf{x}_M) \!-\! \mathbf{m}_0]^{\dag}\mathbf{R}_0^{\!-\!1}[\mathbf{m}_1^{\ast}(\mathbf{x}_M) \!-\! \mathbf{m}_0]\bigg).\label{d_t1}
\end{align}
As per the decision rule in \eqref{decision_rule} and the definitions of the false positive and detection rates, we obtain the desirable results in \eqref{alpha_r} and \eqref{beta_r} after some algebraic manipulations.
\end{IEEEproof}

\begin{figure}[!t]
    \begin{center}
   {\includegraphics[width=3.5in, height=2.9in]{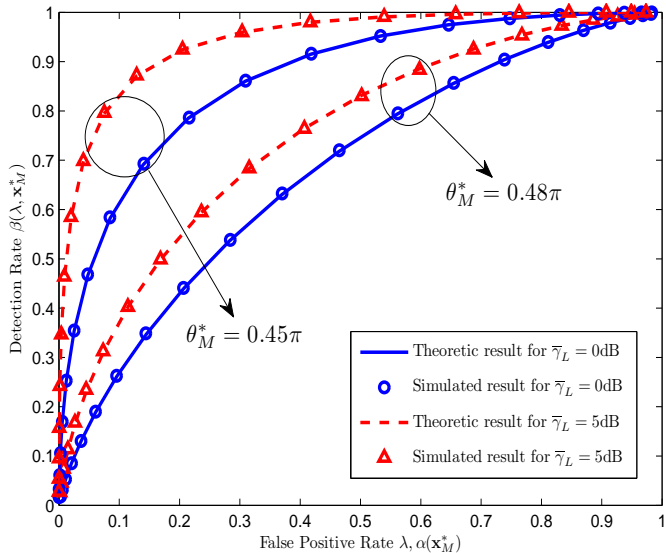}}
    \caption{ROC curves of our LSDS for $N_B = 4$, $N_L=3$, $N_M=3$, $\theta_L = \pi/2$, $K_L = K_M = 1$dB, $\sigma_L^2 = \sigma_M^2 = 0$dB, $P_M = P_M^{\ast}(\mathbf{x}_M^{\ast})$, and $\mathbf{p} = \mathbf{p}^{\ast}(\mathbf{x}_M^{\ast}, P_M^{\ast}(\mathbf{x}_M^{\ast}))$.}\label{fig:fig1}
    \end{center}
\end{figure}

The minimum total error conditioned on a $\mathbf{x}_M$ can be expressed as \cite{barkat2005signal}
\begin{align}
\epsilon (\mathbf{x}_M) = 1 - \beta(\lambda,\mathbf{x}_M) +\alpha(\lambda,\mathbf{x}_M).
\end{align}
We note that the detection performance of the LSDS based on $\mathbf{x}_M^{\ast}$ can be obtained by substituting $\mathbf{x}_M^{\ast}$ into our derived performance metrics. We also note that a decision similar to \eqref{decision_rule} can be obtained for the case where $\mathbf{R}_0 \neq \mathbf{R}_1$. Under this case, the false positive and detection rates cannot be obtained in closed-form expressions since the distribution of the corresponding test statistic is intractable. However, we can utilize a similar methodology presented in \cite{nevat2014distributed} to approximate the distributions of the test statistics in order to obtain the approximations of the false positive and detection rates. Due to the limited space, we left such analysis for further work and we investigate the detection performance of the LSDS for $\mathbf{R}_0 \neq \mathbf{R}_1$ through numerical simulations in the following section.

\section{Numerical Results}

In this section, we first present numerical simulations to verify the accuracy of our provided analysis. We also provide some useful insights on the impact of the SNR of the legitimate channel, the location of the malicious vehicle, number of antennas (i.e., $N_B$, $N_L$, $N_M$), and Rician $K$-factors (i.e., $K_L$, $K_M$) on the detection performance of our LSDS.

In Fig.~\ref{fig:fig1}, we present the Receiver Operating Characteristic (ROC) curve of our LSDS.
In this figure, we first observe that the Monte Carlo simulations precisely match the theoretic results, which confirms our analysis provided in Theorem~1. We also observe that the ROC curves for $\overline{\gamma}_L = 5$dB dominate the ROC curves for $\overline{\gamma}_L = 0$dB. This observation demonstrates that the detection performance of the LSDS increases as the legitimate vehicle's transmit power increases. This is due to the fact that the impact of the channel noise will be relatively suppressed by increasing the transmit power. As expected, we further observe that the ROC curve shifts towards the left-upper corner as $|\theta_M^{\ast}\!-\!\theta_L|$ increases. This demonstrates the necessity to guarantee a minimum distance between the malicious vehicle's claimed location and its true location.

\begin{figure}[!t]
    \begin{center}
   {\includegraphics[width=3.5in, height=2.9in]{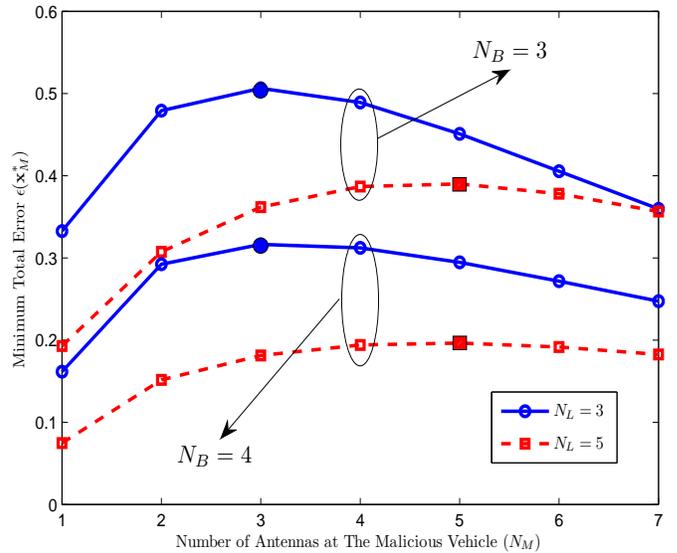}}
    \caption{Minimum total error versus $N_M$ of our LSDS for $\theta_L = \pi/3$, $\theta_M^{\ast} = \pi/4$, $K_L = K_M = 1$dB, $\sigma_L^2 = \sigma_M^2 = 0$dB, $\overline{\gamma}_L = 0$dB, $P_M = P_M^{\ast}(\mathbf{x}_M^{\ast})$, and $\mathbf{p} = \mathbf{p}^{\ast}(\mathbf{x}_M^{\ast}, P_M^{\ast}(\mathbf{x}_M^{\ast}))$.}\label{fig:fig2}
    \end{center}
\end{figure}

\begin{figure}[!t]
    \begin{center}
   {\includegraphics[width=3.5in, height=2.9in]{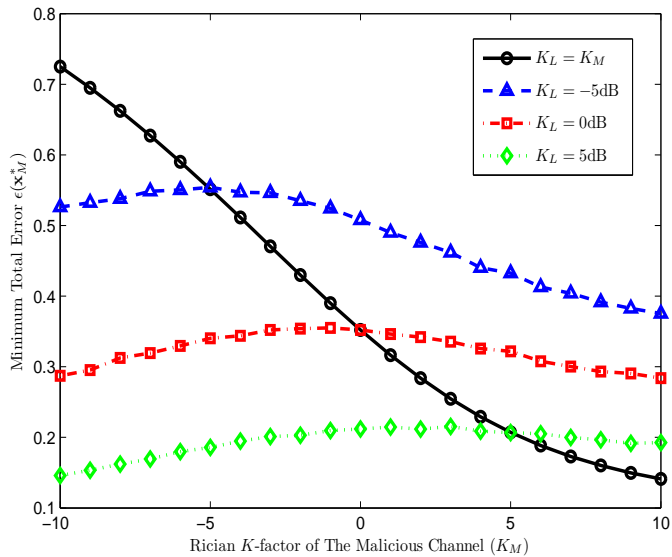}}
    \caption{Minimum total error versus $K_M$ of our LSDS for $N_B = 4$, $N_L = 3$, $N_M = 3$, $\theta_L = \pi/3$, $\theta_M^{\ast} = \pi/4$, $\sigma_L^2 = \sigma_M^2 = 0$dB, $\overline{\gamma}_L = 0$dB, $P_M = P_M^{\ast}(\mathbf{x}_M^{\ast})$, and $\mathbf{p} = \mathbf{p}^{\ast}(\mathbf{x}_M^{\ast}, P_M^{\ast}(\mathbf{x}_M^{\ast}))$.}\label{fig:fig3}
    \end{center}
\end{figure}

In Fig.~\ref{fig:fig2}, we present the minimum total error versus the number of antennas at the malicious vehicle ($N_M$) of our LSDS. As expected, we first observe that the minimum total error decreases as $N_B$ or $N_L$ increases. This is due to the fact that the more antennas the legitimate or the BS is equipped with, the more beamforming gain we can achieve for the legitimate channel. In addition, it is interesting to observe that the minimum total error is maximized when $N_M \!=\! N_L$ for arbitrary $N_B$. This shows that the optimal number of antennas utilized by the malicious vehicle to minimize the detection rate is the same as the number of antennas at the legitimate vehicle. This indicates that if the malicious vehicle is equipped with more antennas than the legitimate vehicle, its attack strategy is to use the same number of antennas as the legitimate vehicle, not to use all of its antennas. In practice, we do not know the number of antennas equipped at the malicious vehicle, but this observation suggests that we can assume $N_M = N_L$ for the attack strategy of the malicious vehicle.

In Fig.~\ref{fig:fig3}, we present the simulated minimum total error versus the Rician $K$-factor of the malicious channel ($K_M$) of our LSDS. As expected, we first confirm that for $K_L = K_M$ the minimum total error decreases as $K_M$ (or $K_L$) increases. In addition, it is interesting to observe that the minimum total error is maximized when $K_M = K_L$ for arbitrary $K_L$. This indicates that the malicious vehicle's attack strategy is to select a true location that is in an environment similar to that of its claimed location (so that $K_M$ is close to $K_L$) to launch location spoofing attacks.

\section{Conclusion}

In this work, we investigated the detection performance of an LSDS for VANETs with a single BS in Rician fading channels. We first determined the malicious vehicle's true location based on a given VANETs scenario and then determined the
optimal transmit power and the optimal directional beamformer for the malicious vehicle to minimize the detection rate. Our analysis first shows that the LSDS performance increases as the Rician $K$-factor of the legitimate channel, the number of antennas at the BS, or the number of antennas at the legitimate vehicle increases. We also obtained a counter-intuitive observation that the malicious vehicle's optimal number of antennas is equal to the legitimate vehicle's number of antennas. Finally, we showed that the Rician $K$-factor of the malicious channel that minimizes the detection rate is identical to the Rician $K$-factor of the legitimate channel.

\section*{Acknowledgments}

This work was funded by The University of New South Wales and Australian Research Council Grant DP120102607.

\end{document}